\documentclass[twocolumn]{aastex631}
\usepackage{amsmath}
\usepackage[english]{babel}
\usepackage[utf8]{inputenc}

 \hypersetup{linkcolor=red,citecolor=blue,filecolor=cyan,urlcolor=magenta}
\shorttitle{Rough Ellipsoid Structure Tools}
\shortauthors{Halder P.}
\graphicspath{{./}{figures/}}

\begin{document}

\title{REST: A java package for crafting realistic cosmic dust particles}

\correspondingauthor{Prithish Halder}
\email{prithishh3@gmail.com, prithish.halder@iiap.res.in}

\author[0000-0002-0786-7307]{Prithish Halder}
\affiliation{Indian Institute of Astrophysics, \\
Koramangala, Bangalore 560034, India}









\begin{abstract}

The overall understanding of cosmic dust particles is mainly inferred from the different Earth-based measurements of interplanetary dust particles and space missions such as Giotto, Stardust and Rosetta. The results from these measurements indicate presence of a wide variety of morphologically significant dust particles. To interpret the light scattering and thermal emission observations arising due to dust in different regions of space, it is necessary to generate computer modelled realistic dust structures of various shape, size, porosity, bulk density, aspect ratio and material inhomogenity. The present work introduces a java package called Rough Ellipsoid Structure Tool (REST), which is a collection of multiple algorithms, that aims to craft realistic rough surface cosmic dust particles from spheres, super-ellipsoids and fractal aggregates depending on the measured bulk-density and porosity. Initially, spheres having $N_d$ dipoles or lattice points are crafted by selecting random material and space seed cells to generate strongly damaged structure, rough surface and poked structure. Similarly, REST generates rough surface super-ellipsoids and poked structure super-ellipsoids from initial super- ellipsoid structures. REST also generates rough fractal aggregates which are fractal aggregates having rough surface irregular grains. REST has been applied to create agglomerated debris, agglomerated debris super-ellipsoids and mixed morphology particles. Finally, the light scattering properties of the respective applied structures are studied to ensure their applicability. REST is a flexible structure tool, which shall be useful to generate various type dust structures that can be applied to study the physical properties of dust in different regions of space. 

\end{abstract}

\keywords{Cosmic dust --- Super-ellipsoids --- Fractal aggregates --- Algorithm --- Software --- Light scattering}


\section{Introduction} \label{sec:intro}

Signatures of dust particles can be found in almost every corner of the visible Universe. The physics of dust plays a crucial role in the understanding of formation and evolution of galaxies as dust particles in a galaxy interact with dust, gas, stars and dark matter under the effect of radiation pressure, magnetic field and gravity \citep{Bekki2015Dust-regulatedParticles}. In most of the studies, the structure of dust is either considered to be spherical or spheroidal in nature. While in reality, dust particles can have different kind of shapes depending on the effect of different forces. The studies related to comet dust and/or atmospheric dust, consider the most complex morphology for the associated dust particles in order to compensate the detailed observations of the structures. Various ground breaking studies such as interplanetary dust particles (IDPs) collected from both the Antarctic ice and the Earth’s stratosphere \citep{Brownlee1985CosmicResearch, Lawler1992CHONHalley, Bradley2007InterplanetaryParticles, Noguchi2015}, the dust particles collected from the Stardust mission \citep{Horz2006ImpactDust} and the micro-physical measurements of dust by the Rosetta/MIDAS and COSIMA instruments \citep{Mannel2019, Guttler2019Synthesis67P/Churyumov-Gerasimenko} prepare a detailed picture of the physical parameters associated with dust particles which are beyond the spherical geometry. Thus, dust particles found in space are not just simple spheres or spheroids, rather they have complex geometries and can be categorised into several groups \emph{viz.}, fractal aggregates, agglomerated debris, rough spheroids, Gaussian random spheres etc. Researchers have often used such complex structures to model the light scattering and thermal emission observations of aerosols, comets, asteroids, protoplanetary disks and debris deisks \citep{Kimura2006LightSpheres, Muinonen2019ScatteringRegoliths, Das2011ModellingParticles, Kolokolova2015PolarizationModel, Zubko2020OnComets, DebRoy2017, Choudhury2020ComparisonDipoles, Halder2018, Halder2021, Aravind2022Optical156P/Russell-LINEAR, Zhuzhulina2022ApertureSimulation, Petrov2022PolarizationParticles}.
Rigorous numerical techniques are required to construct such complex irregular structures. In recent years  material scientists from different fields such as atmospheric physics and nano physics have introduced super-spheroids and/or super-ellipsoids into their study in order to model the observed optical response from such complex structures. Super-ellipsoids are highly generalised geometrical structures whose shape depends on the aspect ratio, north-south and east-west components defined by Equation-\ref{eqn1} in Section-\ref{RSE}. \cite{Lin2018} used super-ellipsoids to model the measured light scattering properties of 25 samples of dust particles (collected from different arid regions on Earth) from the Amsterdam Light Scattering database. On the other hand, \cite{Chatterjee2021PlasmonicDots} used super-ellipsoids to study the variation of plasmonic sensing with changing morphology of nano particles. Application of super-ellipsoids in the study of Astrophysical dust shall provide simpler solutions despite having irregular geometry. 
With the evolution of interpretations of cosmic dust properties from spherical $\rightarrow$ spheroidal $\rightarrow$ irregular, the related light scattering theories have also evolved. For example, the Mie scattering theory that is applicable to single spheres has evolved into T-matrix theory \citep{Waterman1971SymmetryScattering, Mishchenko1996T-matrixReview, Bi2013, Bi2022ComputationMethod} and Discrete Dipole Approximation \citep{Purcell1973ScatteringGrains, Draine1994, Yurkin2011TheLimitations} which are applicable to arbitrary and/or highly irregular dust structures. 

Various applications have emerged to configure and generate ballistic aggregates, ellipsoids, cube, rectangular block, etc., but negligible applications are there to generate rough or irregular geometries. This led us to the development of the package Rough Ellipsoid Structure Tools (REST) that can generate rough surface (RS), poked structures (PS), agglomerated debris (AD) out of ellipsoids and superellipsoids. Apart from the single particle structures REST also generates rough fractal aggregates (RFA) which are fractal aggregates having rough surface irregular grains/monomers from fractal aggregates of spheres (FA). The development of the package, the input and output parameters and algorithm for the different structures are discussed in Section-\ref{sec2}. The physical parameters associated with the structures are explained in Section-\ref{sec3}. Some selected applications of REST and related light scattering results are depicted in Section-\ref{sec4}. 

The instructions related to the basic usage of REST are provided in the online documentation\footnote{\url{https://rest-package.readthedocs.io/}}. REST will become publicly available through the online documentation link in the due course of time. An online request form\footnote{\url{https://form.jotform.com/drprithishhalder/rough-ellipsoid-structure-tools}} related to access of different structures generated using REST is available in the online documentation. The request form provides the same set of input options that are present in REST where, one can choose the different structures and associated physical parameters.
 
\begin{figure*}
    \centering
    \includegraphics[scale=0.28]{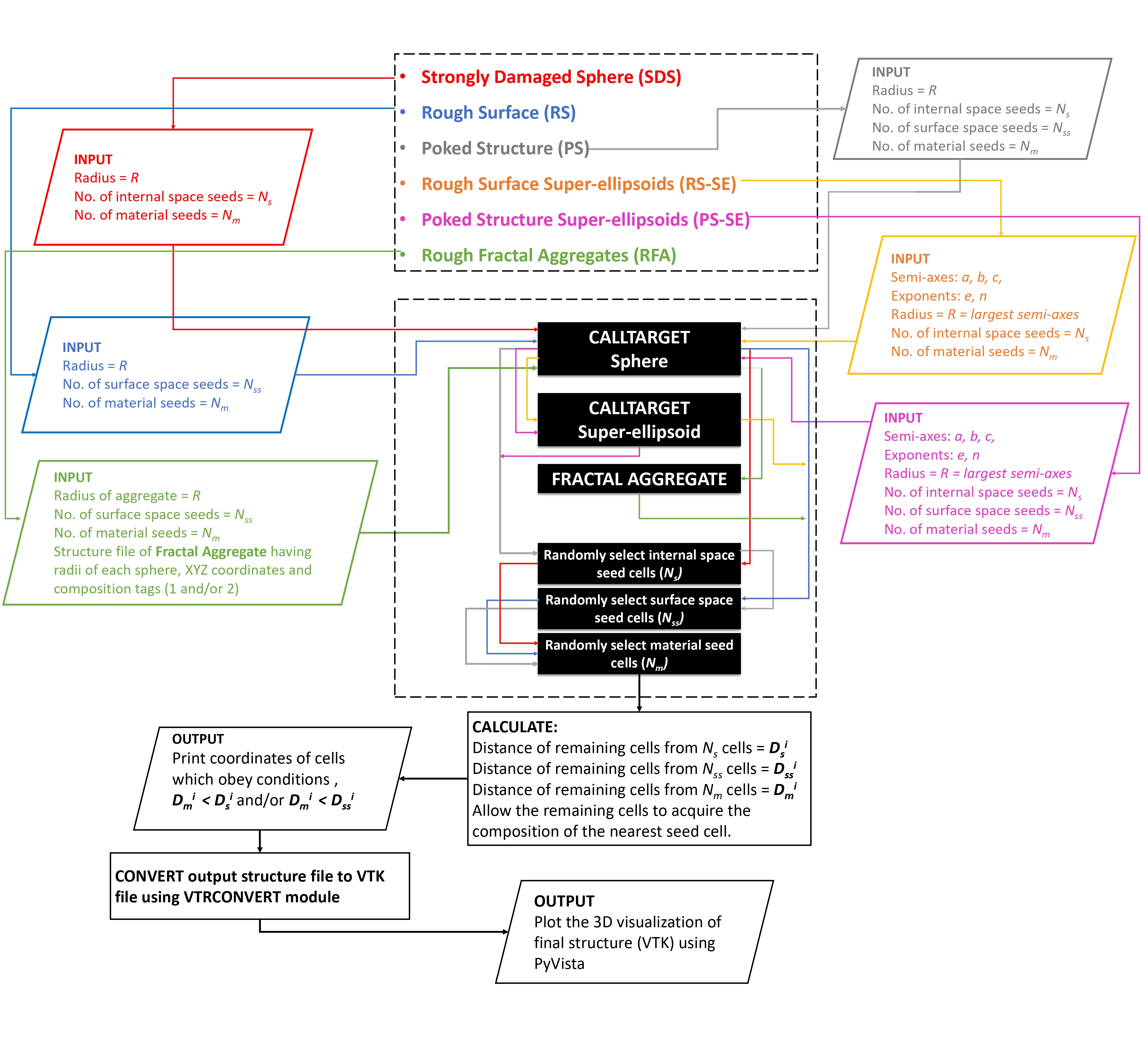}
    \caption{Flow diagram of the package REST}
    \label{fig:algorithm}
\end{figure*}

\section{Rough Ellipsoid Structure Tools (REST)}\label{sec2}
The package REST is a \textsc{Fortran} package wrapped with \textsc{Java} GUI that aims to generate computer models of realistic cosmic dust particles in the form of rough/irregular/broken/porous structures. REST is composed of three main parts: (1) \textsc{Java} GUI takes the input parameters, (2) \textsc{Fortran} modules performs the calculations to craft roughness or breaking the initial spherical, super-ellipsoidal or fractal aggregate structures and (3) \textsc{Python3} scripts plots the respective structures using VTK and PyVista\footnote{\url{https://www.pyvista.org/}} \citep{Sullivan2019PyVista:VTK} python modules. REST uses the CALLTARGET module from DDSCAT \citep{Draine1994} to generate the initial spherical and super-ellipsoidal structures.

The different input and output parameters of REST are shown below:

The input parameters are\-
\begin{itemize}
    \item Type of structure
    \begin{itemize}
        \item Strongly Damaged Sphere (SDS)
        \item Rough surface (RS) 
        \item Poked Structure (PS)
        \item Rough Surface Super-ellipsoids (RS-SE)
        \item Poked structure Super-ellipsoids (PS-SE)
        \item Rough fractal aggregates (RFA)
    \end{itemize}
    \item Radius for spherical initial structure or circumscribed sphere \emph{R$_d$}
    \item Initial number of dipoles \emph{N$_{d}$}
    \item Semi-axes \emph{a, b \& c} of the initial super-ellipsoidal structure
    \item East-west exponent \emph{e} of the initial super-ellipsoidal structure
    \item North-south exponent \emph{n} of the initial super-ellipsoidal structure
    \item Radii and coordinate points of each sphere of fractal aggregate (\textit{r, X,Y,Z})
    \item Number of material seed cells \emph{N$_{m}$}
    \item Number of space seed cells \emph{N$_{s}$}
    \item Number of surface space seed cells \emph{N$_{ss}$}
    \item Thickness of the surface layer \emph{t}
    \item Final target file name
\end{itemize}

The output parameters are,
\begin{itemize}
    \item Initial \& final \textsc{DDSCAT} structure target file
    \item Initial \& final \textsc{VTK} structure file
    \item Structure plot png file.
\end{itemize}

The Java GUI asks the user to choose the type of structure and enter the respective input parameters such as size (radius in case of SDS, RS and PS structures; semi-axes and exponents in case of RS-SE and PS-SE structures), number of material seed cells, number of internal space seed cells and/or number of surface seed cells. In case of RFA structures, the applications asks the user to choose the structure file of a FA structure having radii, coordinates and composition tag of each sphere/monomer. After entering the input parameters, the user can proceed for calculation by clicking the “Calculate” button in the Java GUI. In each case, the base structure is a sphere. For SDS, RS and PS, the radius of the initial sphere is $R_d$. While, in case of RS-SE and PS-SE the radius of the initial sphere is equal to the largest semi-axes. In case of RFA structures, the radius of the initial sphere is equivalent to the radius of the FA structure. The initial spherical and super-ellipsoidal structures are generated using the CALLTARGET module considering the radius coming out from the input parameters discussed above.
The initial and the final structure files are composed of seven columns, one for the dipole number (JA), three for the dipole coordinates (IX, IY, IZ) and the remaining three for the respective composition tags [ICOMP(x,y,z)]. The SDS, RS and PS structures discussed bellow are generated following \cite{Zubko2006DDAStructure} and the structures RS-SE, PS-SE, and RFA are generated using algorithms discussed in this study for the first time. Figure-\ref{fig:algorithm} shows the flow diagram of the package showing workflow of the six FORTRAN modules explained below:




\begin{figure*}
    \centering
    \includegraphics[scale=0.26]{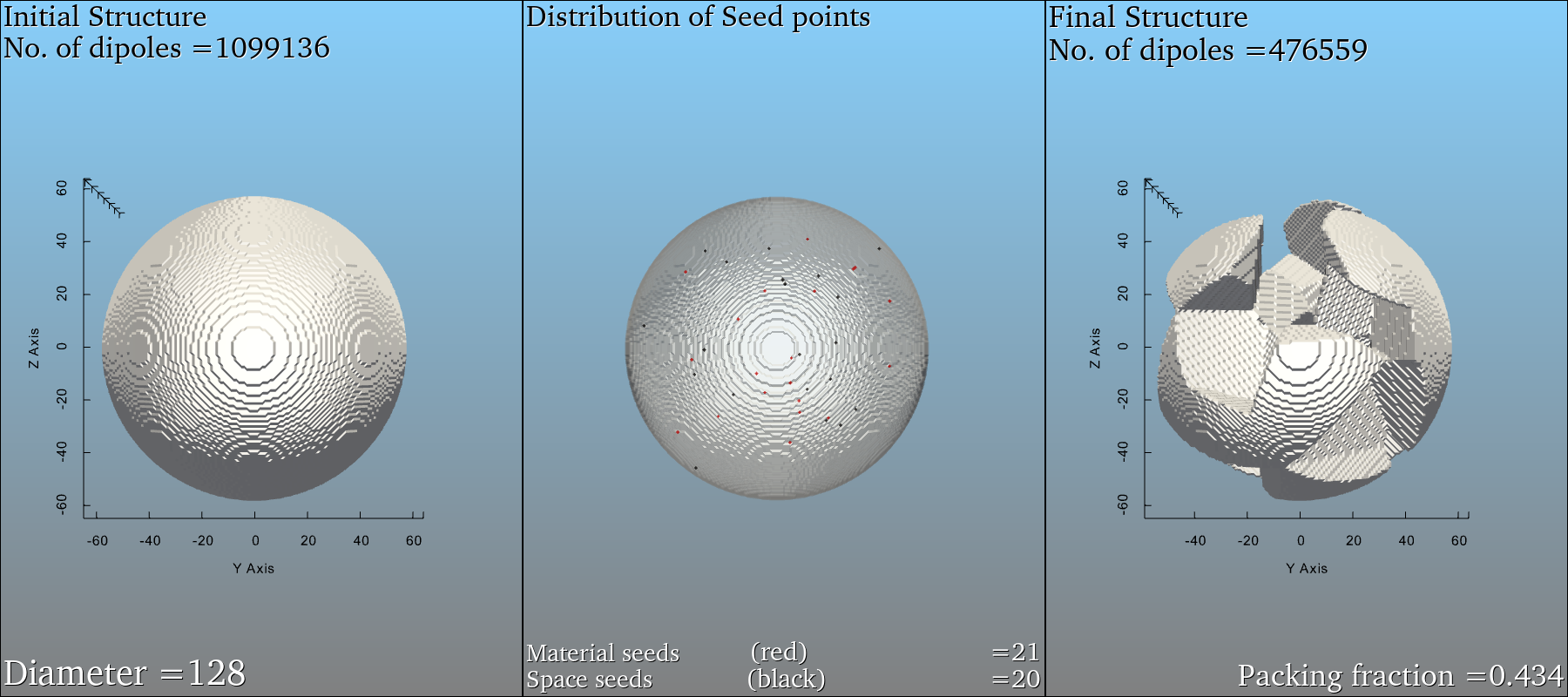}
    \caption{3D Visulization of SDS (default) structure using REST (PyVista module)}
    \label{fig:sds_default}
\end{figure*}

\begin{figure*}
    \centering
    \includegraphics[scale=0.26]{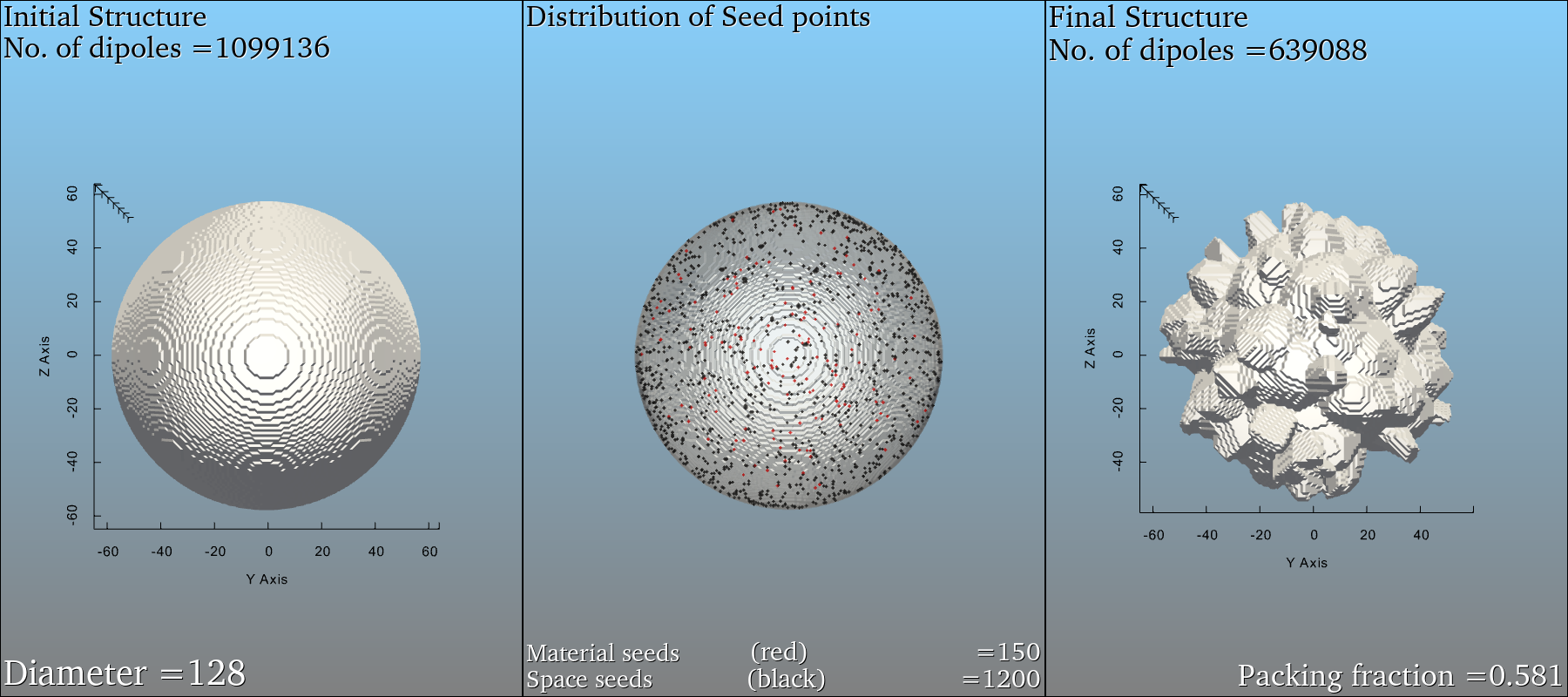}
    \caption{3D Visulization of RS structure using REST (PyVista module)}
    \label{fig:rs}
\end{figure*}

\begin{figure*}
    \centering
    \includegraphics[scale=0.26]{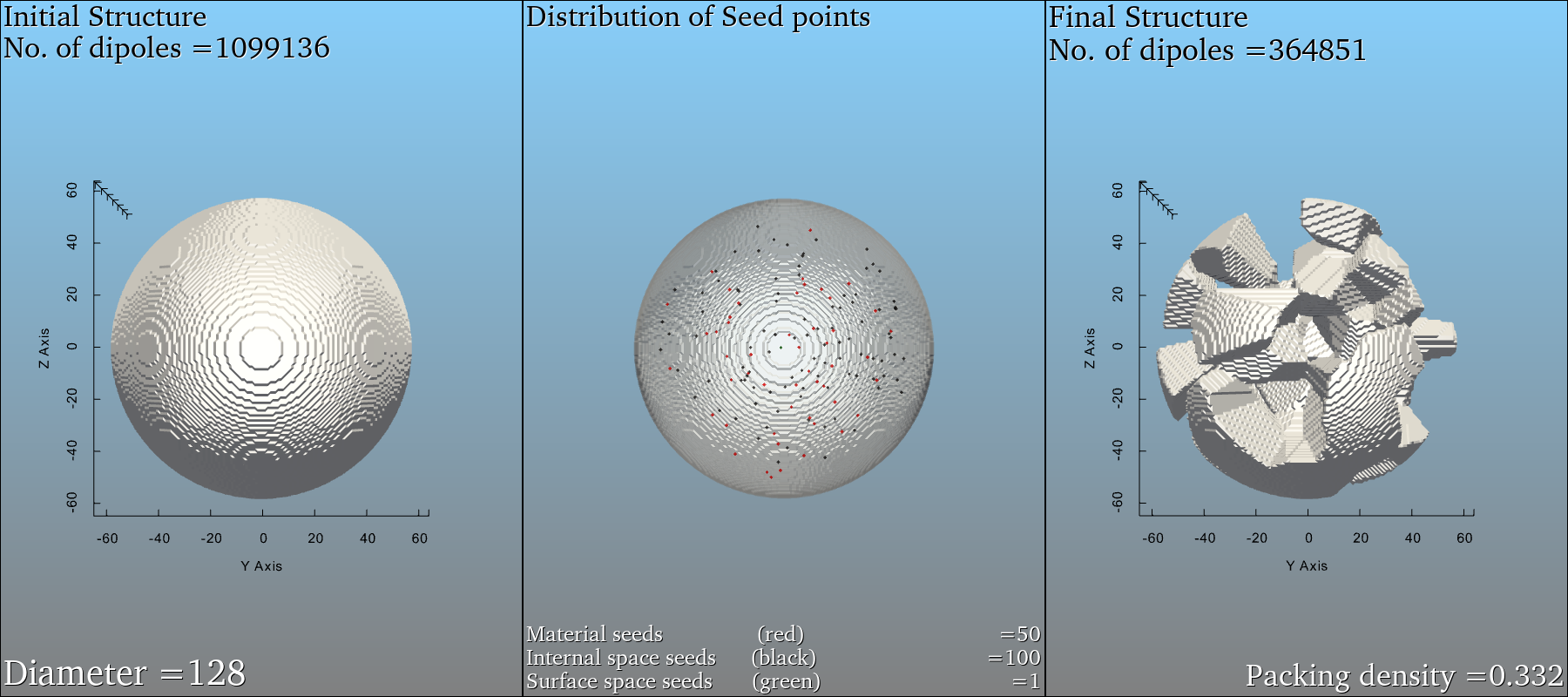}
    \caption{3D Visulization of PS structures using REST (PyVista module)}
    \label{fig:ps}
\end{figure*}

\subsection{Strongly Damaged Sphere (SDS)}
\label{subsubsection:sds}
The SDS is a kind of structure that has been broken from certain portion of its volume, yet retaining a partial spherical shape. To generate SDS structure REST follows the steps stated below:

\begin{enumerate}
    \item Generate initial spherical structure file \texttt{target.out} having $N_{d}$ dipoles and radius $R$ (in number of dipoles) using \texttt{CALLTARGET} module.

    
    
    \item Randomly choose $N_{m}$ number of material seed cells from the $N_{d}$ dipoles present in the \texttt{target.out} file.
    \item Randomly choose $N_{s}$ number of space seed cells from the $N_{d}$ dipoles present in the \texttt{target.out} file.
    \item Measure the distance $D_{m}^{i,j}$ between the $j$th material seed cell and $i$th dipole of the base structure, where $j$ = 1 to $N_{m}$ and $i$ = 1 to $N_{d}$.
    \item Measure the distance $D_{s}^{i,k}$ between the $k$th space seed cell and $i$th dipole of the base structure, where $k$ = 1 to $N_{s}$ and $i$ = 1 to $N_{d}$.
    \item Print those dipoles for which, $D_{m}^{i}$ $<$ $D_{s}^{i}$ in the final structure file.
\end{enumerate}
The development of SDS structure is shown in Figure-\ref{fig:sds_default}. REST has two options for selecting seed cells for SDS structure, (i) \texttt{Default seeds} and (ii) \texttt{Custom seeds}.

\subsection{Rough surface (RS)}\label{RS-PS}
The RS structures are spherical structures having rough surface. Unlike, SDS, the space seed cells in RS structures are selected on the surface of the base structure having certain amount of thickness. 
To generate RS structure REST follows the algorithm stated below:
\begin{enumerate}
    \item Generate initial spherical structure file \texttt{target.out} having $N_{d}$ dipoles and radius $R$ (in number of dipoles) using \texttt{CALLTARGET} module.
    \item Randomly choose $N_{m}$ number of material seed cells from the $N_{d}$ dipoles present in the \texttt{target.out} file.
    \item Randomly choose $N_{ss}$ number of surface space seed cells from the $N_{d}$ dipoles present in the \texttt{target.out} file. The surface thickness should be $t$ times $r$.
    \item Measure the distance $D_{m}^{i,j}$ between the $j$th material seed cell and $i$th dipole of the base structure, where $j$ = 1 to $N_{m}$ and $i$ = 1 to $N_{d}$.
    \item Measure the distance $D_{ss}^{i,l}$ between the $l$th surface space seed cell and $i$th dipole of the base structure, where $l$ = 1 to $N_{ss}$ and $i$ = 1 to $N_{d}$.

    \item Print those dipoles for which, $D_{m}^{i}$ $<$ $D_{ss}^{i}$ in the final structure file.
\end{enumerate}
The development of RS structure is shown in Figure-\ref{fig:rs}.

\subsection{Poked Structures (PS)}\label{ps}
The PS are structures which are poked from inside as well as from the surface. For PS structures, two types of space seeds are selected. One for the internal space and the other for the surface space.
To generate PS structure REST follows the algorithm stated below:
\begin{enumerate}
    \item Generate initial spherical structure file \texttt{target.out} having $N_{d}$ dipoles and radius $R$ (in number of dipoles) using \texttt{CALLTARGET} module.

    \item Randomly choose $N_{m}$ number of material seed cells from the $N_{d}$ dipoles present in the \texttt{target.out} file.
    
    \item Randomly choose $N_{is}$ number of internal space seed cells from the $N_{d}$ dipoles present in the \texttt{target.out} file.
        
    \item Randomly choose $N_{ss}$ number of surface space seed cells from the $N_{d}$ dipoles present in the \texttt{target.out} file. The surface thickness should be $t$ times $r$.
    
    \item Measure the distance $D_{m}^{i,j}$ between the $j$th material seed cell and $i$th dipole of the base structure, where $j$ = 1 to $N_{m}$ and $i$ = 1 to $N_{d}$.
    \item Measure the distance $D_{is}^{i,k}$ between the $k$th internal space seed cell and $i$th dipole of the base structure, where $k$ = 1 to $N_{is}$ and $i$ = 1 to $N_{d}$.
    \item Measure the distance $D_{ss}^{i,l}$ between the $l$th surface space seed cell and $i$th dipole of the base structure, where $l$ = 1 to $N_{ss}$ and $i$ = 1 to $N_{d}$.
   \item Print those dipoles for which, $D_{m}^{i}$ $<$ $D_{is}^{i}$ and $D_{m}^{i}$ $<$ $D_{ss}^{i}$ in the final structure file.

\end{enumerate}
The development of PS structure is shown in Figure-\ref{fig:ps}.

The PS structure option asks the user to choose from homogeneous and inhomogeneous morphology. In case of homogeneous morphology REST takes material seeds for one kind of material, while in case of inhomogeneous morphology, REST takes two kinds of material seeds $N_{m1}$ \& $N_{m2}$. Thus the overall material seeds, $N_{m}$ = $N_{m1}$ + $N_{m2}$.

\begin{figure*}
    \centering
    \includegraphics[scale=0.26]{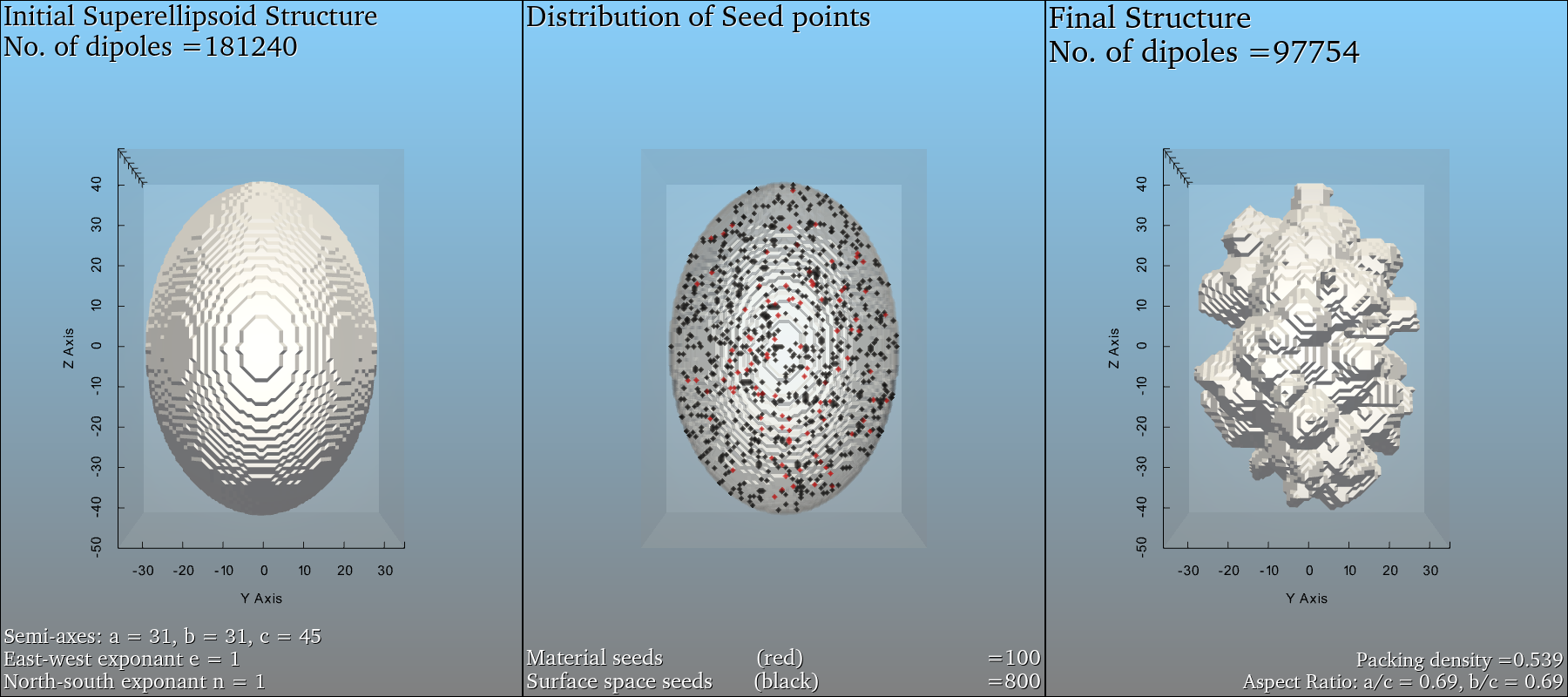}
    \caption{3D Visulization of RS-SE structure using REST (PyVista module)}
    \label{fig:rs_se}
\end{figure*}

\begin{figure*}[htp]
    \centering
    \includegraphics[scale=0.26]{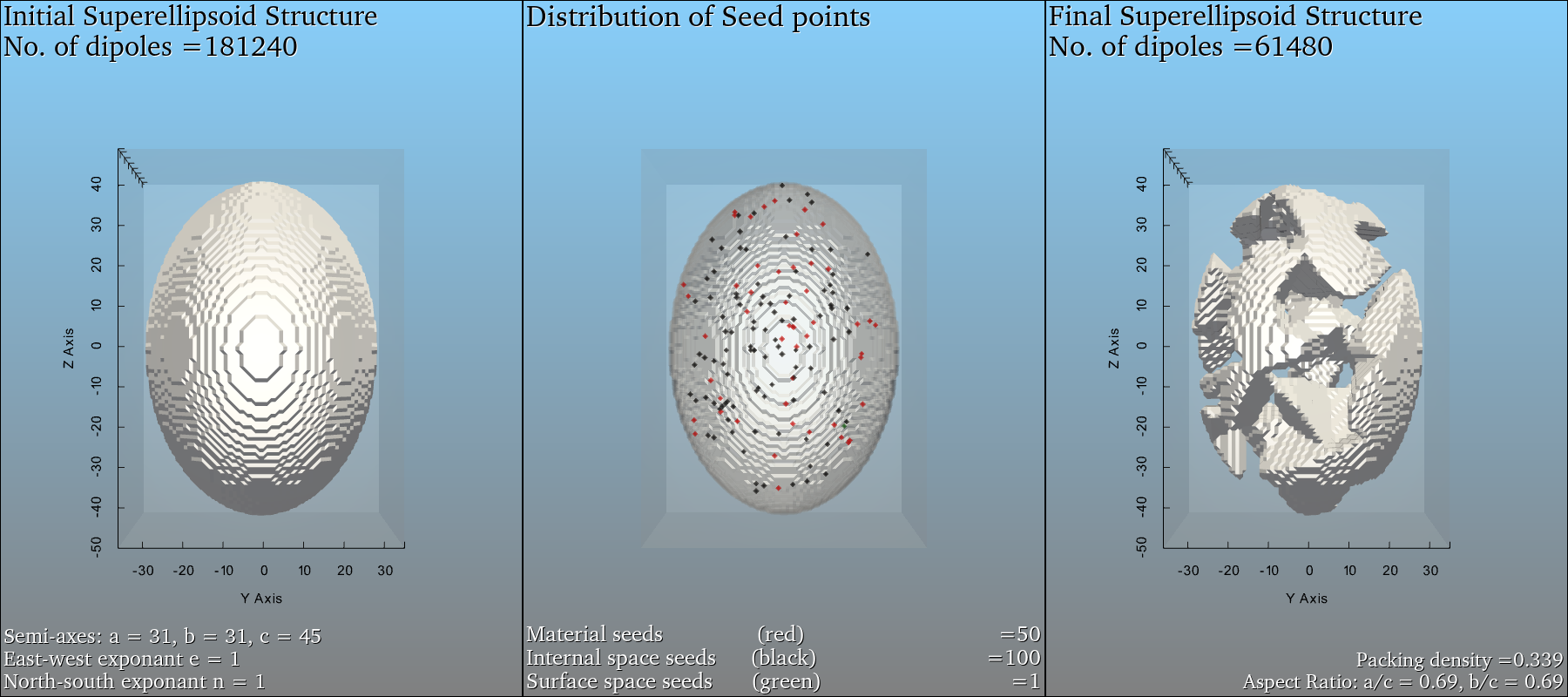}
    \caption{3D Visulization of PS-SE structure using REST (PyVista module)}
    \label{fig:ps_se}
\end{figure*}

\subsection{Rough Super-ellipsoids}{\label{RSE}}
Rough Super-ellipsoids are super-ellipsoidal structures having rough or irregular surface or volume morphology. Super-ellipsoids (SE) are generalised ellipsoidal structures which are defined by the following equation \citep{Faux1979ComputationalManufacture, Barr1981SuperquadricsTransformations, Bi2018AssessingSpace}:
\begin{equation}\label{eqn1}
   \Bigg[\bigg(\frac{x}{a}\bigg)^{2/e} + \bigg(\frac{y}{b}\bigg)^{2/e}\Bigg]^{e/n} + \Bigg[\bigg(\frac{z}{c}\bigg)\Bigg]^{2/n} = 1
\end{equation}

where $a$, $b$ and $c$ are the semi-axes of the SE structure along the $x$, $y$ and $z$ directions in Cartesian coordinates. The parameters $e$ and $n$ in Equation \ref{eqn1} are the East-west and North-south exponents of the SE structure (also called roundness parameters). The roundness parameters determine the structural variation for constant aspect ratio ($a$/$b$ or $b$/$c$).

REST uses SE as base structures to create Rough Surface (RS-SE) and Poked Structure (PS-SE).
To generate RS-SE/PS-SE structures REST uses the following algorithm:
\begin{enumerate}
    \item Generate the initial super-ellipsoidal structure having $N_{d}$ dipoles and semi-axes $a$, $b$ and $c$ (in number of dipoles) using the \texttt{SUPELLIPS} function from \texttt{CALLTARGET} module.
    \item Create a circumscribed sphere having radius $R$ = longer semi-axis of initial super-ellipsoidal structure using the \texttt{ELLIPSOID} function from \texttt{CALLTARGET} module.
    \item Randomly choose $N_{ss}$ surface space seed cells from the boundary where the surface of SE meets that of the circumscribed sphere.
    \item Randomly choose $N_{m}$ material seed cells from within the volume of the SE.
    \item Measure the distance $D_{m}^{i,j}$ between the $j$th material seed cell and $i$th dipole of the base structure, where $j$ = 1 to $N_{m}$ and $i$ = 1 to $N_{d}$.
    \item Measure the distance $D_{ss}^{i,k}$ between the $k$th surface space seed cell and $i$th dipole of the base structure, where $k$ = 1 to $N_{ss}$ and $i$ = 1 to $N_{d}$.
    \item Print those dipoles for which, $D_{m}^{i}$ $<$ $D_{ss}^{i}$ in the final structure file.
    \item For PS-SE structure Randomly choose $N_{is}$ internal space seed cells from the boundary where the surface of SE meets that of the circumscribed sphere along with above mentioned steps.
    \item Measure the distance $D_{is}^{i,k}$ between the $k$th internal space seed cell and $i$th dipole of the base structure, where $k$ = 1 to $N_{is}$ and $i$ = 1 to $N_{d}$.
    \item Print those dipoles for which, $D_{m}^{i}$ $<$ $D_{is}^{i}$ and $D_{m}^{i}$ $<$ $D_{ss}^{i}$ in the final structure file.
\end{enumerate}
The development of RS-SE and PS-SE structures are shown in Figures-\ref{fig:rs_se} \& \ref{fig:ps_se} respectively.

\begin{figure*}[!htp]
    \centering
    \includegraphics[scale=0.26]{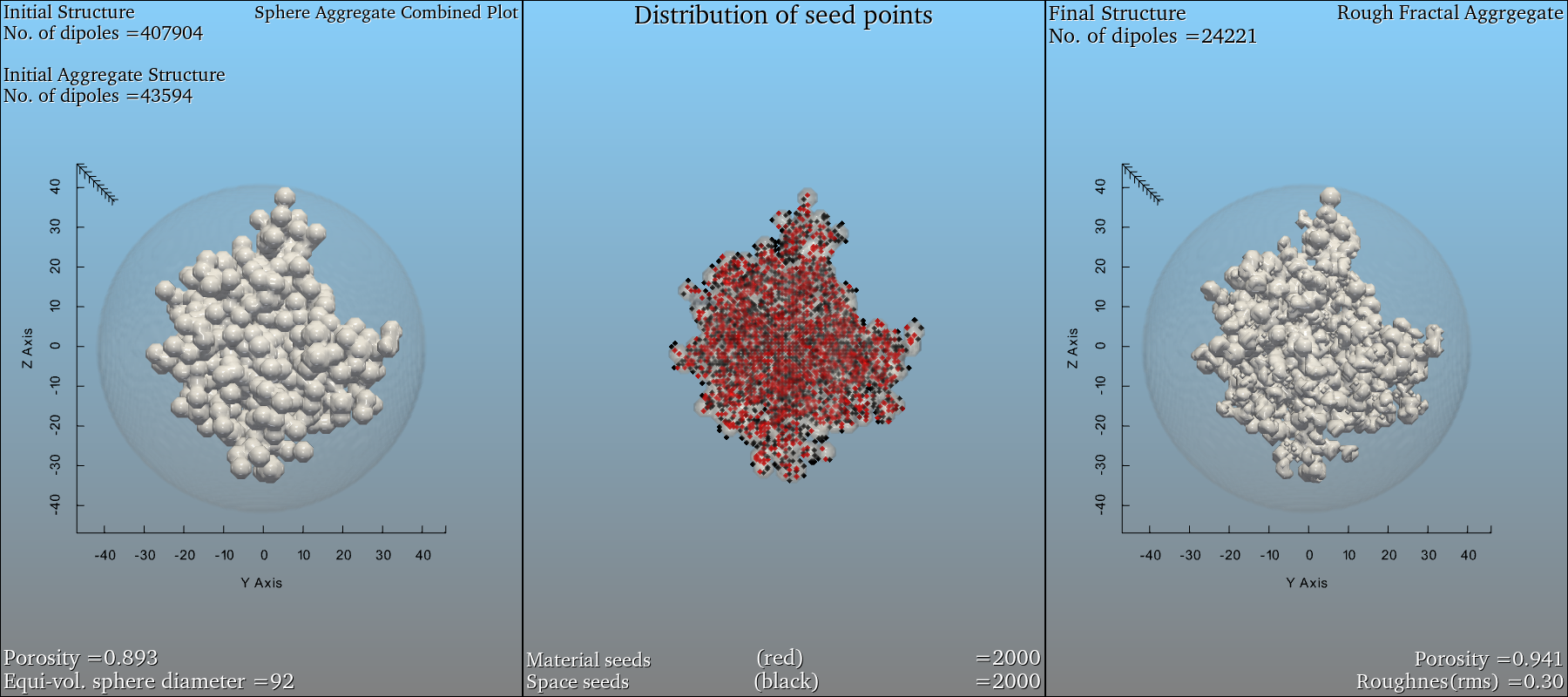}
    \caption{3D Visulization of RFA structure using REST (PyVista module)}
    \label{fig:rfa_1}
\end{figure*}

\subsection{Rough Fractal Aggregates (RFA)}
The RFA are fractal aggregates made up of cluster of irregular rough surface grains/monomers/sub-units. The the IDPs obtained from the Earth's stratosphere indicate presence of FA of irregular rough surface grains. REST converts the smooth spherical surface of each spheres of a particular fractal aggregate/cluster of spheres into rough and irregular surface grains or monomers. The following algorithm allows REST to craft RFA structures from FA structures made up of spheres:

\begin{enumerate}
    \item Browse and select the structure file of a fractal aggregate/cluster of spheres having following format ($i$, $R$, $X$ $Y$ $Z$, \emph{mtag}, \emph{mtag}), where $i$ is the sphere number, $R$ is the $i^{th}$ sphere radius (in $\mu$m), $X$, $Y$, $Z$ are the coordinates of each sphere (monomer) and the \emph{mtag} is the composition tag.
    
    \item Multiply each coordinate and radii with an integer scale factor $n$.
    
    \item Measure the distance ($d_{mono}$) of each monomer from the centre ($0$,$0$,$0$).

    \item Create the initial sphere having $N_{d}$ dipoles and radius $R_d$ = \texttt{maximum}($d_{mono}$) (in number of dipoles) from the centre of the initial sphere.
    
    \item Randomly choose $N_{ss}$ surface seed cells on the surface of the FA inside the initial sphere. 
    
    \item Randomly choose $N_{m}$ material seed cells inside the surface of the FA.
    
    \item Measure the distance $D_{m}^{i,j}$ between the $j$th material seed cell and $i$th dipole of the base structure, where $j$ = 1 to $N_{m}$ and $i$ = 1 to $N_{d}$.
    
    \item Measure the distance $D_{ss}^{i,k}$ between the $k$th surface space seed cell and $i$th dipole of the base structure, where $k$ = 1 to $N_{ss}$ and $i$ = 1 to $N_{d}$.
    
    \item Print those dipoles for which, $D_{m}^{i}$ $<$ $D_{ss}^{i}$ in the final RFA structure file.

\end{enumerate}

The development of RFA structure is shown in Figure-\ref{fig:rfa_1}. 

The RFA structure generated using REST retains the fractal nature from the initial FA or cluster of sphere. The roughness crafted on the final structure induces certain amount of porosity which can be controlled by changing the ratio of material and space seed cells ($N_{m}$:$N_{ss}$). The RFA structure is a unique structure and holds very close resemblance in terms structural morphology with the structures of IDPs obtained from the Earth's stratosphere and the \emph{Stardust} mission.

\begin{figure*}[!htp]
    \centering
    \includegraphics[scale=0.26]{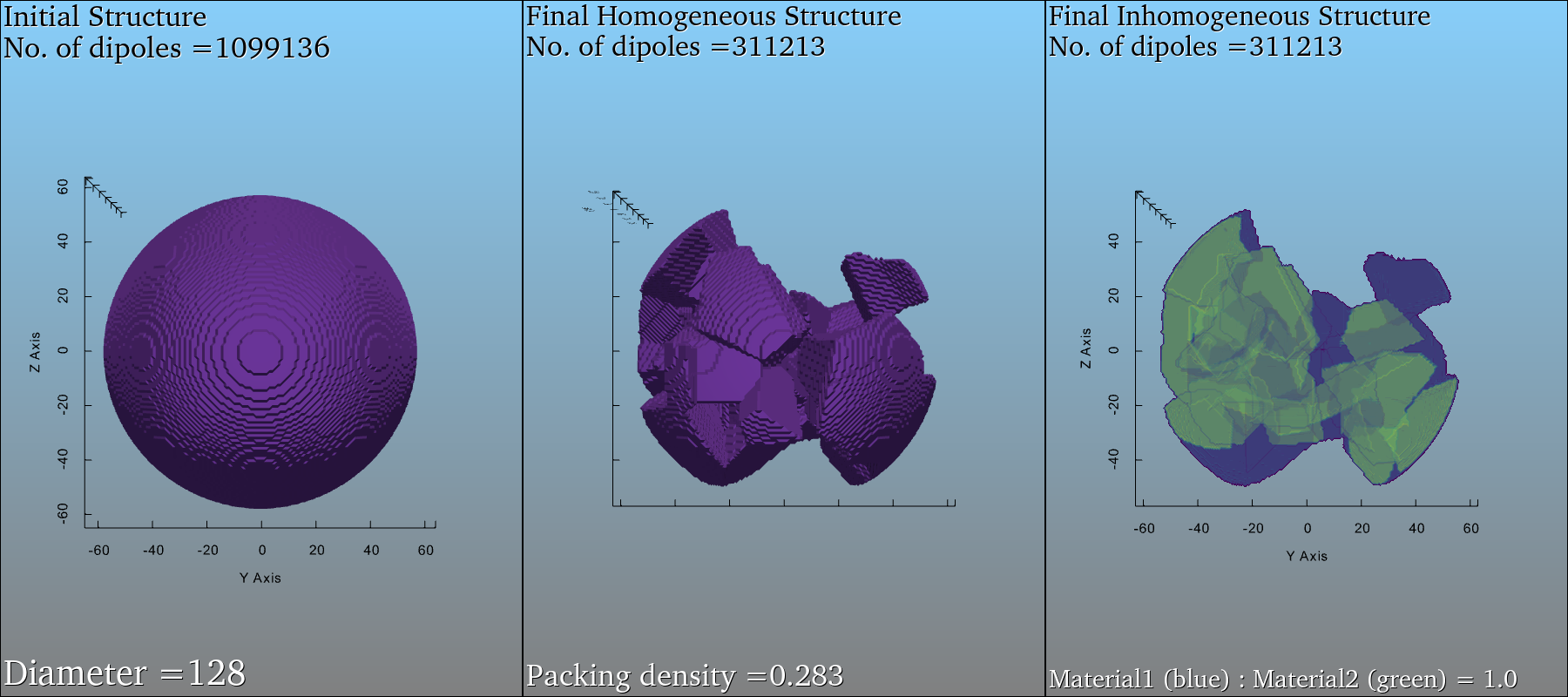}
    \caption{3D Visulization of inhomogeneous PS structures using REST (PyVista module)}
    \label{fig:ps_inh}
\end{figure*}

\begin{figure*}[!htp]
    \centering
    \includegraphics[scale=0.26]{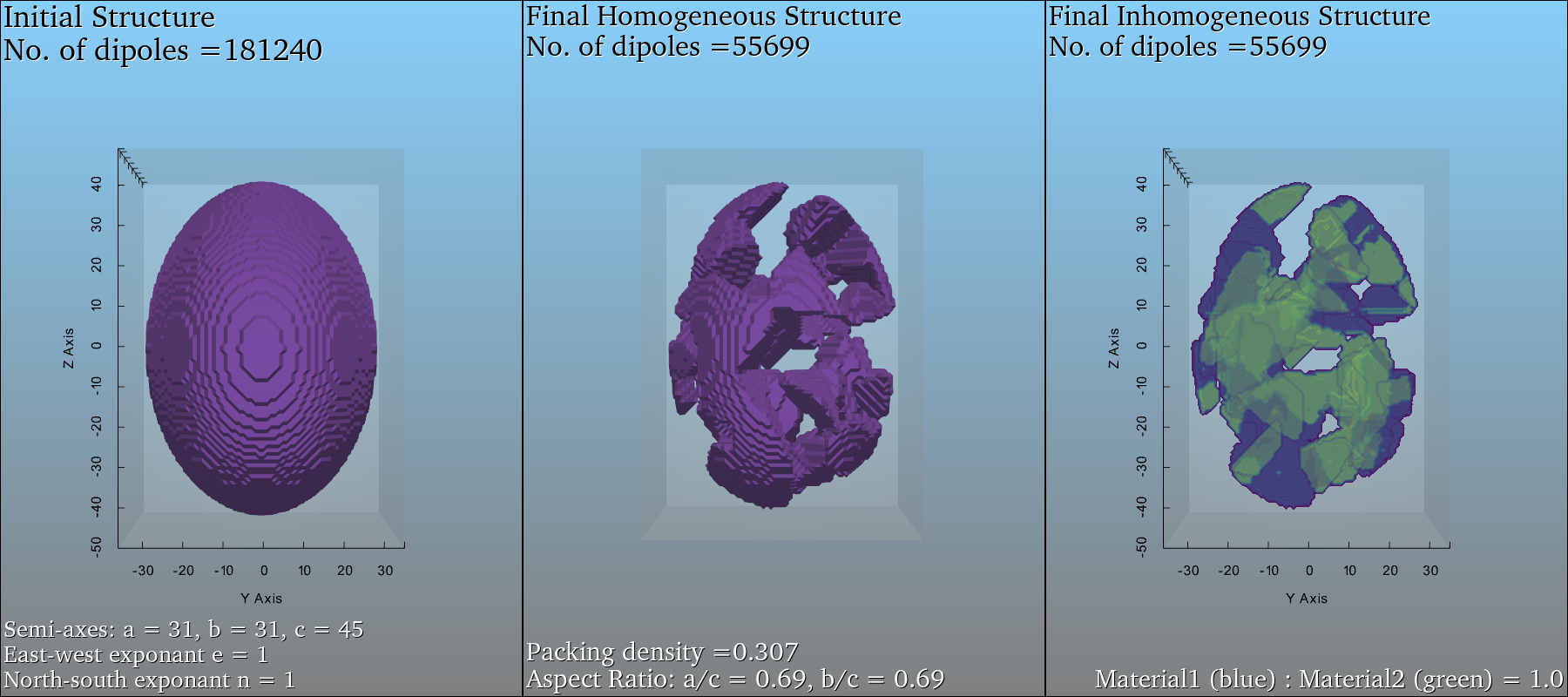}
    \caption{3D Visulization of inhomogeneous PS-SE structure using REST (PyVista module)}
    \label{fig:ps-se_inh}
\end{figure*}

\begin{figure*}[!htp]
    \centering
    \includegraphics[scale=0.26]{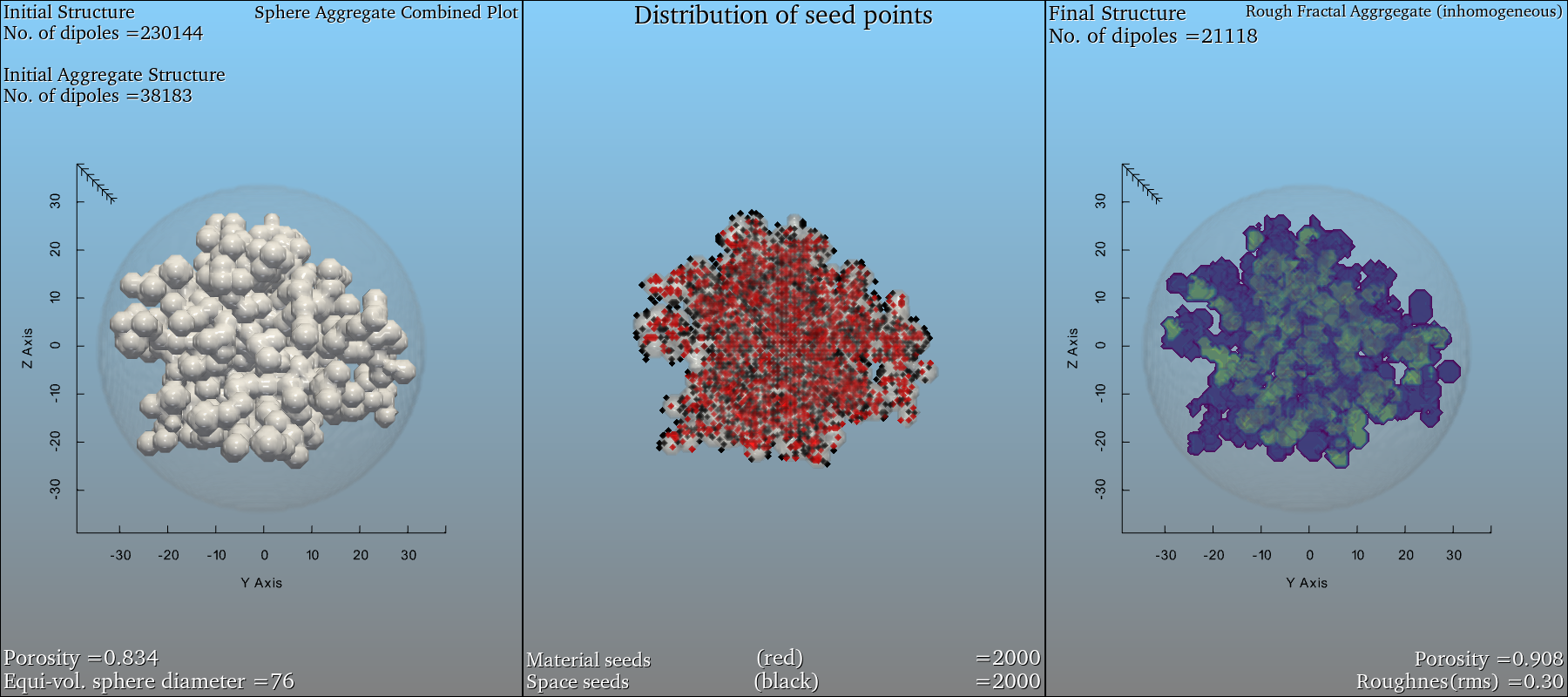}
    \caption{3D Visulization of inhomogeneous RFA structure using REST (PyVista module)}
    \label{fig:rfa_inh}
\end{figure*} 

\subsection{Inhomogeneous structures}\label{inh}
REST allows users to choose from homogeneous and inhomogeneous morphology. In case
of homogeneous morphology REST takes material seeds
for one kind of material as explained in the earlier subsections, while in case of inhomogeneous morphology, it takes two kinds of material seeds $N_{m1}$ \& $N_{m2}$. Thus, the overall material seeds, $N_{m}$ = $N_{m1}$ + $N_{m2}$. The inhomogeneous option is available for structure modules PS, PS-SE and RFA. Figures-\ref{fig:ps_inh}, \ref{fig:ps-se_inh} \& \ref{fig:rfa_inh} shows the construction of respective inhomogeneous structures.

\section{Physical parameters}\label{sec3}
The development of the structures discussed in Section-\ref{sec2}, deals with geometry of the dipole matrix generated using the CALLTARGET function. The unit of all the parameters in the algorithm are based on the number of dipoles. Hence, in this section the associated physical parameters of the structures and their units are discussed.

\subsection{Radius}
In REST the radius $R_d$ or the semi-axes $a$, $b$ and $c$, must be chosen in terms of number of dipoles from the origin to the surface.
If $R_d$=32 (no. of dipoles from the origin to the surface), then diameter = 64 and $N_{d}$ = 137,376. In light scattering calculations the size parameter $X$ of a particle is used to solve the scattering problem. $X$ is defined by the following equation,
\begin{equation}
  X = \frac{2\pi R}{\lambda}
\end{equation}

where $R$, is the radius (in physical units) and $\lambda$ is the wavelength of incident radiation (in physical units). Follow Table-\ref{tab:table1} to choose the value of radius $R_d$ (in no. of dipoles) \citep{Zubko2020OnComets} or semi-axes.

\begin{table}
    \centering
    \begin{tabular}{|c|c|c|c|}
    \hline

       X & Radius ($R_d$) & Diameter & $N_{d}$\\
    \hline
        X $<$ 16 & 32 & 64 & 137,376\\
        X = 16-32 & 64 & 128 & 1,099,136\\
        X $>$ 32 & 128 & 256 & 8,783,848\\
    \hline
    \end{tabular}
    \caption{Chart to choose the radius in number of dipoles according to the required size parameter (X).}
    \label{tab:table1}
\end{table}

For example, \newline if $R$ = 1.0$\mu$m, $\lambda$ = 0.450$\mu$m then, \newline \[X = \frac{2\pi R}{\lambda} = \frac{2\times 3.142 \times 1.0}{0.450} = 13.96 < 16\]
As, $X$ $<$ 16, the radius in dipoles $R_d$ = 32.

This procedure for selection of the radius is for better coverage of mesh or dipoles so that the DDSCAT calculation accuracy is better. One can always use radius of their choice, but calculation accuracy may decrease if the values decrease from those shown in Table-\ref{tab:table1}.

\subsection{Semi-axes}
In SE options, REST takes the semi-axes $a$, $b$ and $c$ as input for the size or dimension of the initial SE structure. For particles with size parameter $X$ $<$ 16, $N_{d}$ of the final structure should be within 20,000 to 40,000 for better accuracy. In case of super-spheroid structure, $a$=$b$ and $a/c$ is the aspect ratio.

\begin{figure*}
    \centering
    \includegraphics[scale=0.55]{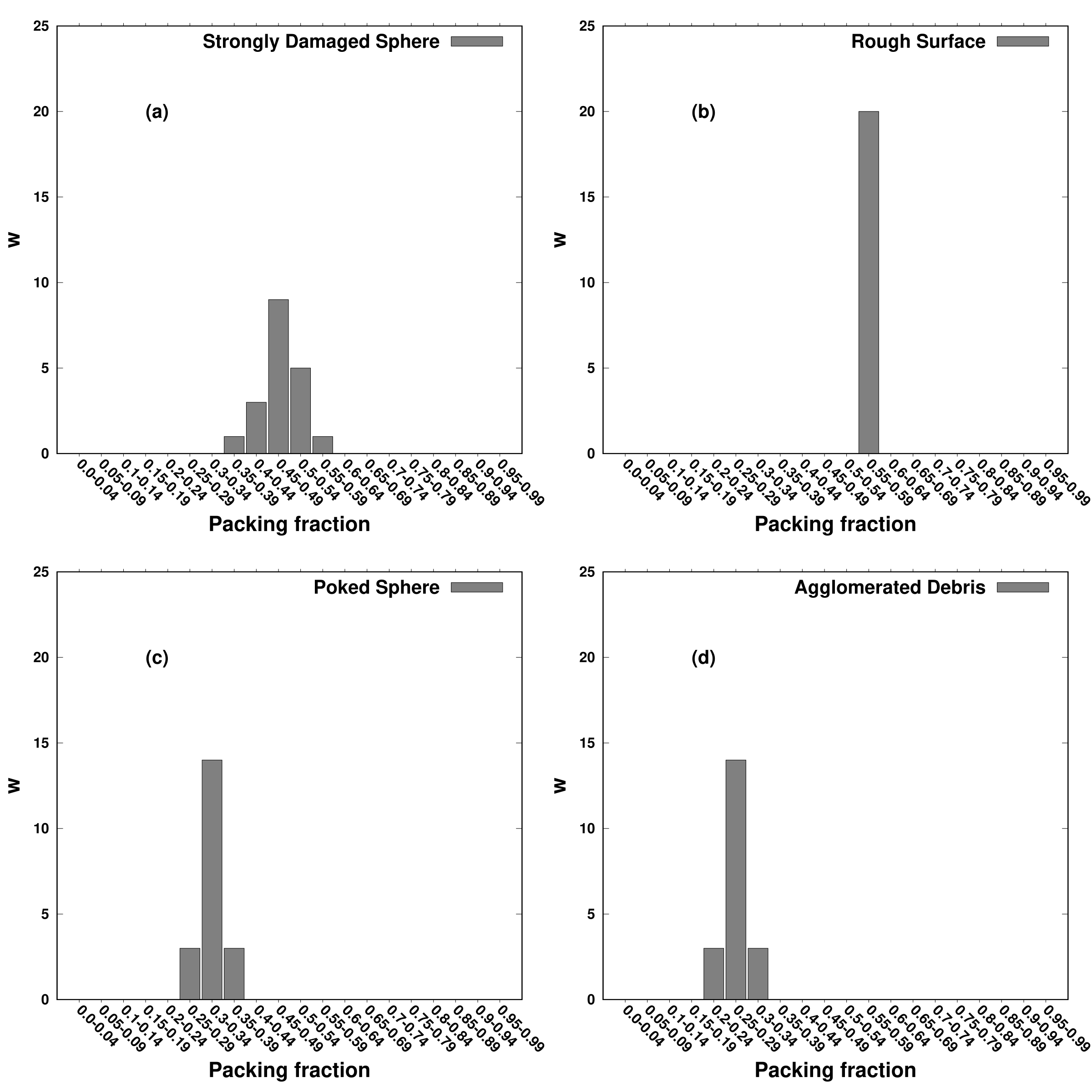}
    \caption{Packing fraction of different structures generated using REST}
    \label{fig:pack_frac}
\end{figure*}

\subsection{Bulk-density}
The bulk density $\rho_{b}$ of a material is defined by the following equation,
\begin{equation}
    \rho_{f} = \frac{\rho_{b}}{\rho}
\end{equation}

where, $\rho_{f}$ and $\rho$ are the packing fraction and true density of the material, respectively. Packing fraction $\rho_{f}$ of a material is the fraction of space occupied by the atoms/the unit cells, in this case its the dipoles.

In REST, the value of $\rho_{f}$ is determined by the following equation,
\begin{equation}
\begin{split}
    \rho_{f} & = \frac{\texttt{Volume of final structure}}{\texttt{Volume of initial structure}} = \frac{V_{f}}{V_{i}} \\
    & = \frac{N_{d}(\texttt{final})\times d^{3}}{N_{d}(\texttt{initial})\times d^{3}} = \frac{N_{d}(\texttt{final})}{N_{d}(\texttt{initial})}
\end{split}
\end{equation}

where, $N_{d}(\texttt{initial})$ and $N_{d}(\texttt{final})$ are the total number of dipoles in the initial and the final structure respectively while $d$ is the dipole spacing. In DDSCAT the volume of a structure is determined by the number of dipoles $N_{d}$ and the dipole spacing $d$,
\begin{equation}
    V = N_{d}\times d^{3}
\end{equation}

Thus, for a RS particle having radius $r$ = 32, $N_{d}$(\texttt{initial}) = 137,376, $N_{d}$(\texttt{final}) = 71,813, the packing fraction is, \newline
\[\rho_{f} = \frac{N_{d}(\texttt{final})}{N_{d}(\texttt{initial})} = \frac{71,813}{137,376} = 0.523\]

The study of comet dust particles collected from the stardust mission reveals the bulk density in the range 0.3 - 3.0 \texttt{gm/cm$^{3}$} with high porous particles having bulk density as low as 0.3 \texttt{gm/cm$^{3}$} and low porous/solid particles having bulk density as high as 3.0 \texttt{gm/cm$^{3}$} \citep{Horz2006ImpactDust}. The value of $\rho_{f}$ for the different structures obtained using REST are shown in Figure-\ref{fig:pack_frac}(a-d).

\subsection{Porosity}
Porosity is one of the significant physical parameters for the RFA structures. The degree of porosity is defined by the ratio of total number of space seed cells in the entire volume circumscribing the RFA structure by the total volume of the circumscribing sphere. Thus, we have,
\[\texttt{Volume of initial structure} (V_i) = N_d(\texttt{initial})\times d^3\]
\[\texttt{Volume of final structure} (V_f)   = N_d(\texttt{final})\times d^3\]

The total volume of space seed cells is given by,

\[V_T = [N_d(\texttt{final}) - N_d(\texttt{initial})]\times d^{3}\]

Therefore, the degree of porosity is,

\begin{equation}
\begin{split}
    P & = \frac{V_T}{V_f} = \frac{[N_d(\texttt{final}) - N_d(\texttt{initial})]\times d^{3}}{N_{d}(\texttt{final})\times d^{3}} \\
    & = \frac{[N_d(\texttt{final}) - N_d(\texttt{initial})]}{N_{d}(\texttt{final})}
\end{split}
\end{equation}

\subsection{Roughness}
In this study, the surface roughness of each monomer/grain of a RFA particle is defined as the mean deviation of the surface normal of a RFA grain or monomer from the initial spherical grain of the initial 
fractal aggregate/cluster of spheres. If $D_{i}$ is the distance of the $i$th dipole or lattice point on the $j$th RFA monomer/grain from the centre of the $j$th RFA monomer/grain, then the roughness on the $j$th monomer/grain is defined the root mean square equation,

\begin{equation}
    R_{rms}^{j} = \bigg\{\frac{1}{N_{j}}\sum_{n=1}^{N_{j}} (D_{i} - \langle\ D_{i} \rangle\ )^{2} \bigg\}^{1/2}
\end{equation}
 
\section{Applications}\label{sec4}
In this section, some of the applications of REST that can be employed to create visually realistic cosmic dust particles has been discussed. AD, AD super-ellipsoids (AD-SE) and RFA-AD mixed morphology particles are generated using REST. The respective particles are used to simulate light scattering using DDSCAT to calculate the the degree of linear polarization for changing phase angle and size parameter considering different material composition. The light scattering parameters are defined by the phase matrix which represents far-field transformation of Stoke's parameters of the incident light ($I_i$, $Q_i$, $U_i$, $V_i$) to that of the scattered light ($I_s$, $Q_s$, $U_s$, $V_s$).   This phase matrix is given by \citep{Bohren1998AbsorptionParticles}:
\begin{equation}
\left( \begin{array}{cccc}
I_s(\theta)  \\
Q_s(\theta)  \\
U_s(\theta)  \\
V_s(\theta)  \end{array} \right)= \frac{1}{k^{2}d^{2}} \left( \begin{array}{cccc}
    S_{11}(\theta) & S_{12}(\theta) & S_{13}(\theta) & S_{14}(\theta) \\
    S_{21}(\theta) & S_{22}(\theta) & S_{23}(\theta) & S_{24}(\theta) \\
    S_{31}(\theta) & S_{32}(\theta) & S_{33}(\theta) & S_{34}(\theta) \\
    S_{41}(\theta) & S_{42}(\theta) & S_{43}(\theta) & S_{44}(\theta) \end{array} \right)\left( \begin{array}{cccc}
I_i(\theta)  \\
Q_i(\theta)  \\
U_i(\theta)  \\
V_i(\theta)  \end{array} \right)
\end{equation}
where, $k$ is the wave-number and $d$ is the distance between the scatterer and the observer and $S_{ij}$ represents the scattering matrix elements. Angle $\theta$ is the \emph{Scattering angle}, $\theta$ = [0$^{\circ}$,180$^{\circ}$]. 

The degree of linear polarization explained below to denote the light scattering behaviour of the respective particles is defined by \(-S_{12}/S_{11}\).

\begin{figure*}
    \centering
    \includegraphics[scale=0.26]{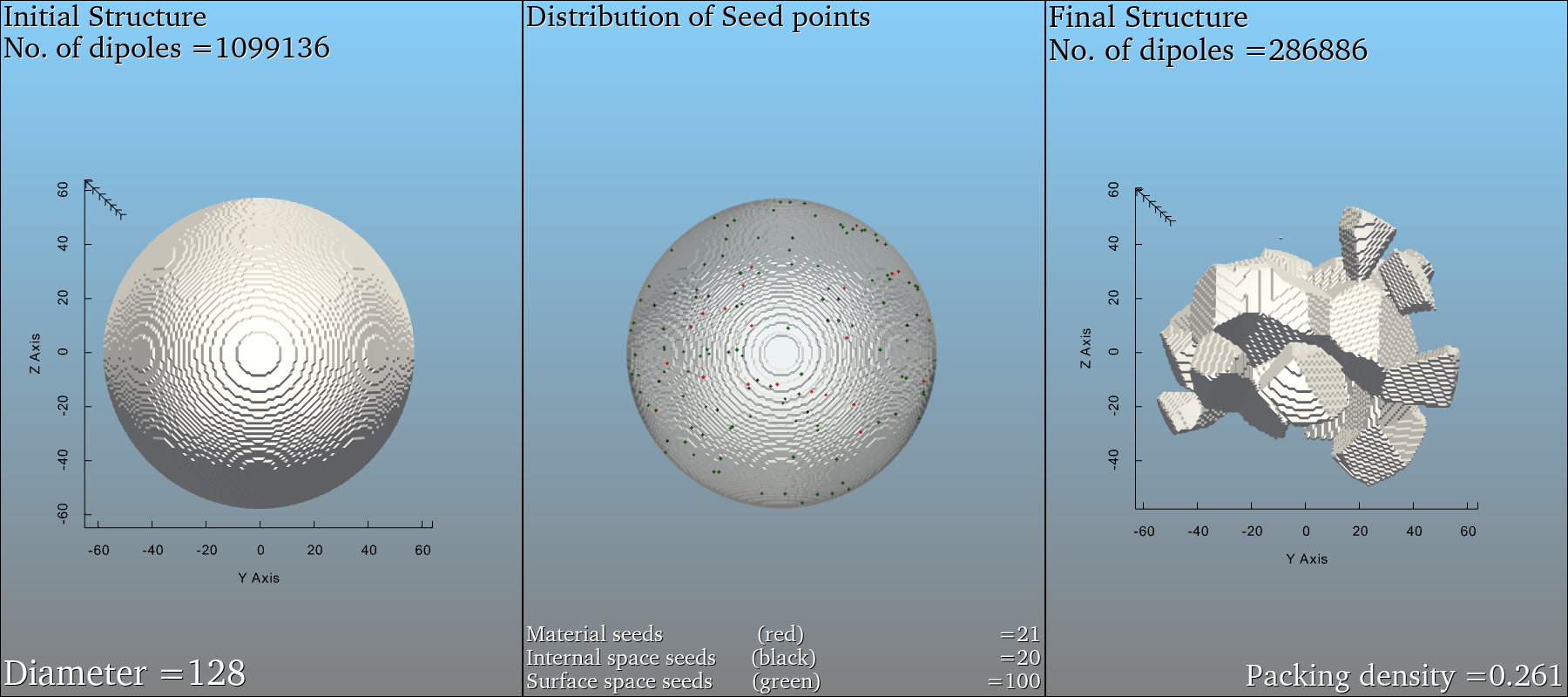}
    \caption{3D Visulization of AD structure using REST (PyVista module)}
    \label{fig:ad}
\end{figure*}

\begin{figure*}
    \centering
    \includegraphics[scale=0.26]{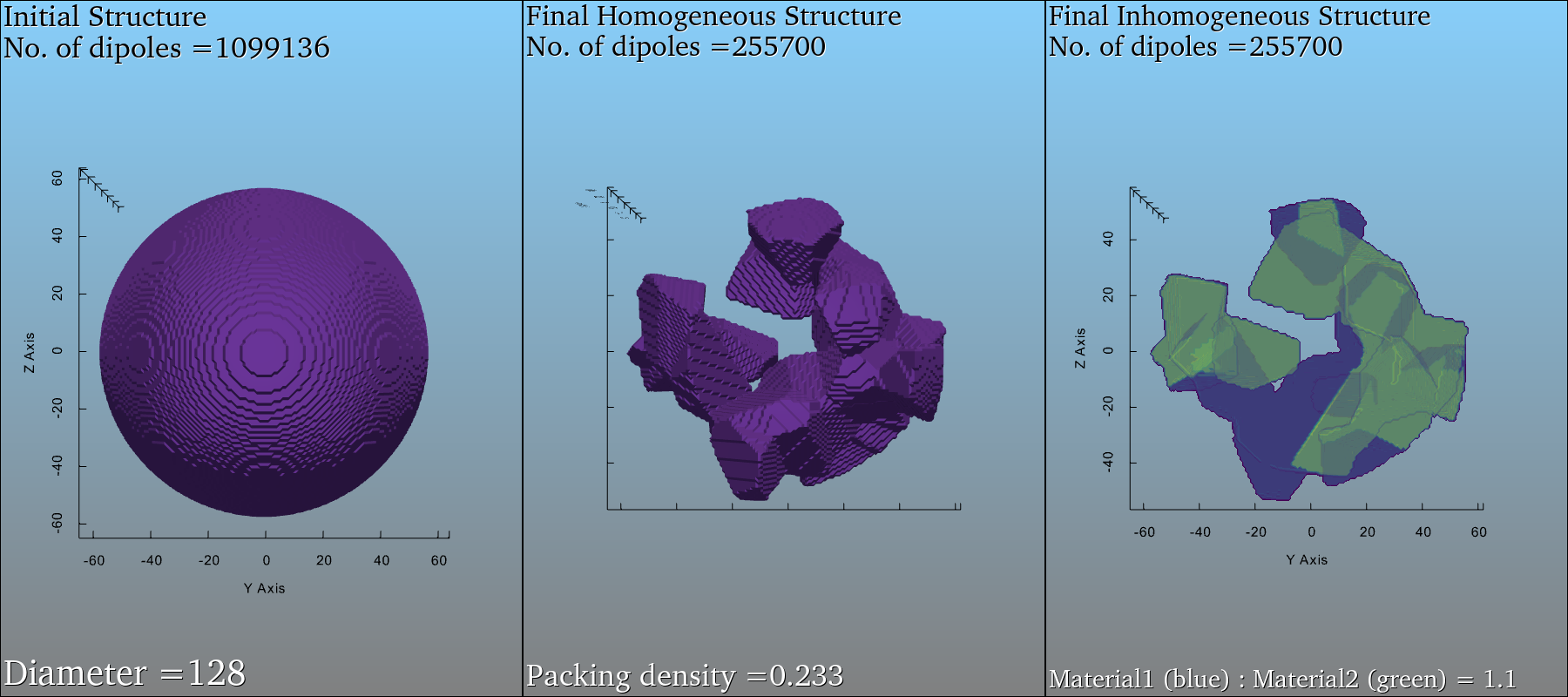}
    \caption{3D Visulization of inhomogeneous AD structure using REST (PyVista module)}
    \label{fig:ad-inh}
\end{figure*}

\subsection{Agglomerated Debris (AD)}
The AD particles are one of the most widely used model cosmic dust particles by different researchers to model the light scattering properties of comet dust, asteroid dust and dust in debris disks. The AD particles have highly irregular morphology with shape and bulk density resembling those of the solid or low porous group of particles found in comets. These particles require relatively lesser number of free parameters to simulate light scattering by cosmic dust. The AD particles are generated using following parameters considering the PS algorithm:
\begin{itemize}
    
    \item $R_d$ (in number of dipoles) = 64
    \item $N_m$ = 21
    \item $N_{is}$ = 20
    \item $N_{ss}$ = 100
    \item $t$ = 1\%

\end{itemize}

\begin{figure*}
    \centering
    \includegraphics[scale=0.28]{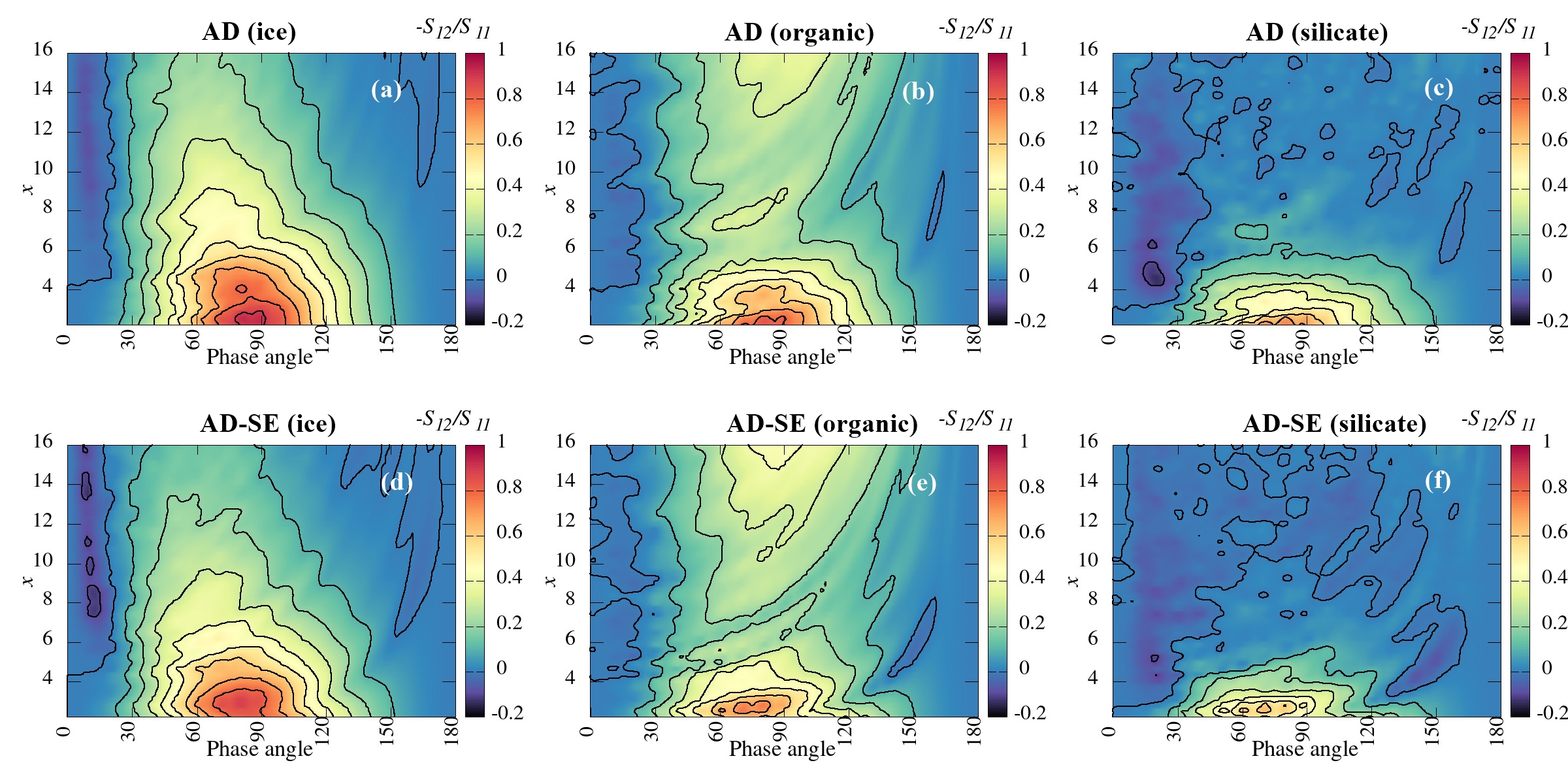}
    \caption{Variation of the degree of linear polarization with phase angle for $X$ = 2 to 16 for AD structures having composition of \emph{ice} (a), \emph{organic} (b) and \emph{silicate} (c) and for AD-SE structures having composition of \emph{ice} (d), \emph{organic} (e) and \emph{silicate} (f)}
    \label{fig:ad_adse}
\end{figure*}

Figure-\ref{fig:ad} shows the AD structure formed using above parameters. The initial spherical structure has $N_d$ = 1099136 and the final AD structure has $N_d$ = 286886 with packing fraction $\rho_{f}$ = 0.261. Figures-\ref{fig:ad_adse}(a-c) shows the variation of the degree of linear polarization ($-S_{12}/S_{11}$) with increasing phase angle ($0^{\circ} - 180^{\circ}$) over a size parameter ($x = 2{\pi}R/\lambda$) rang 2 to 16 for material compositions of ice, organic and silicates. The results are similar to those obtained for AD particles in \cite{Zubko2006DDAStructure}. Figure-\ref{fig:ad-inh} shows the inhomogeneous AD particles generated using REST considering the PS and inhomogeneous algorithm discussed in Section-2.6.

\begin{figure*}
    \centering
    \includegraphics[scale=0.26]{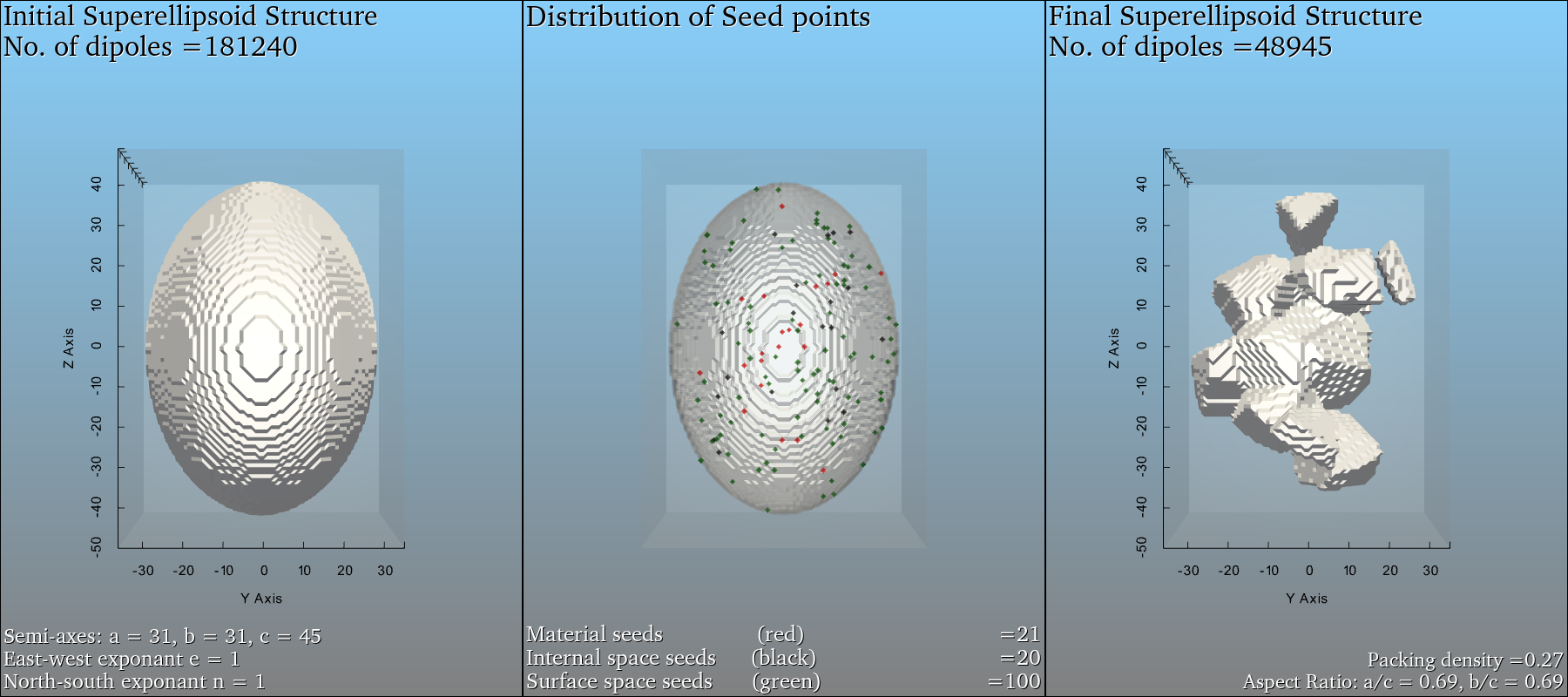}
    \caption{3D Visulization of AD-SE structure using REST (PyVista module)}
    \label{fig:ad_se}
\end{figure*}

\begin{figure*}
    \centering
    \includegraphics[scale=0.26]{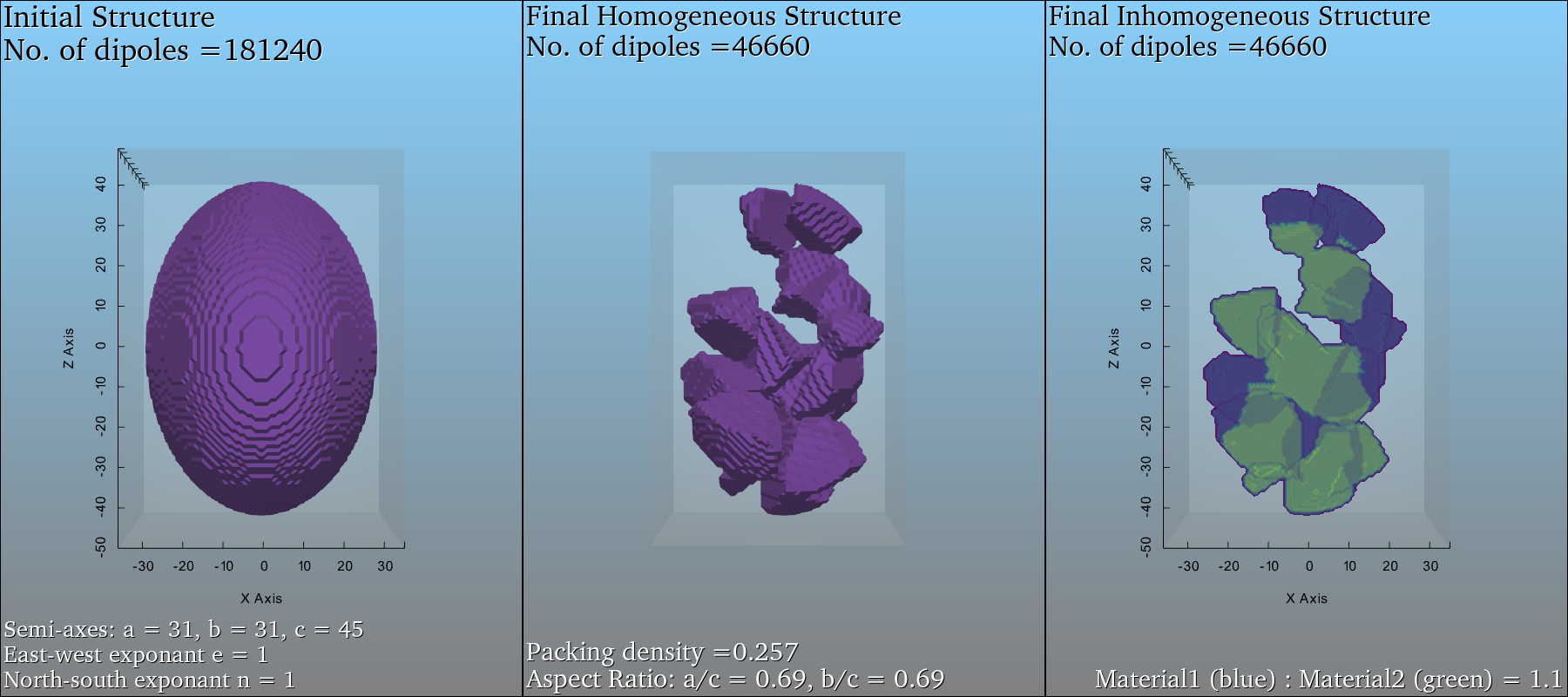}
    \caption{3D Visulization of inhomogeneous AD-SE structure using REST (PyVista module)}
    \label{fig:ad-se-inh}
\end{figure*}

\subsection{Agglomerated Debris Super-ellipsoids (AD-SE)}
The AD-SE are agglomerated debris particles confined within a super-ellipsoidal volume and hence one can study the effect of aspect ratio. The dust particles present in circumstellar or protoplanetary environment posses certain order of aspect ratio and are believed to be oriented under the effect of stellar magnetic field. To study the dust in such environments, the AD-SE particles shall provide better interpretation of the observed polarization and extinction. The AD-SE particles are generated using following parameters considering the PS-SE algorithm:
\begin{itemize}
    
    \item Semi-axes: $a$ = 31, $b$ = 31, $c$ = 45
    \item Exponants: $n$ = 1, $e$ = 1
    \item $N_m$ = 21
    \item $N_{is}$ = 20
    \item $N_{ss}$ = 100

\end{itemize}

Figure-\ref{fig:ad_se} shows the AD-SE structure formed using above parameters. The initial SE structure has $N_d$ = 181240 and the final AD-SE structure has $N_d$ = 48945 with packing fraction $\rho_{f}$ = 0.27 and aspect ratio: $a/c$ = $b/c$ = 0.69. Figures-\ref{fig:ad_adse}(d-f) shows the variation of the degree of linear polarization ($-S_{12}/S_{11}$) with increasing phase angle ($0^{\circ} - 180^{\circ}$) over a size parameter ($x = 2{\pi}R/\lambda$) range 2 to 16 for material compositions of ice, organic and silicates. The scattered light in case of AD-SE is more depolarized compared to those obtained for AD particles for all the three compositions. Figure-\ref{fig:ad-se-inh} shows the inhomogeneous AD-SE particles generated using REST considering the PS-SE and inhomogeneous algorithm discussed in Section-\ref{inh}.

\begin{figure*}
    \centering
    \includegraphics[scale=0.22]{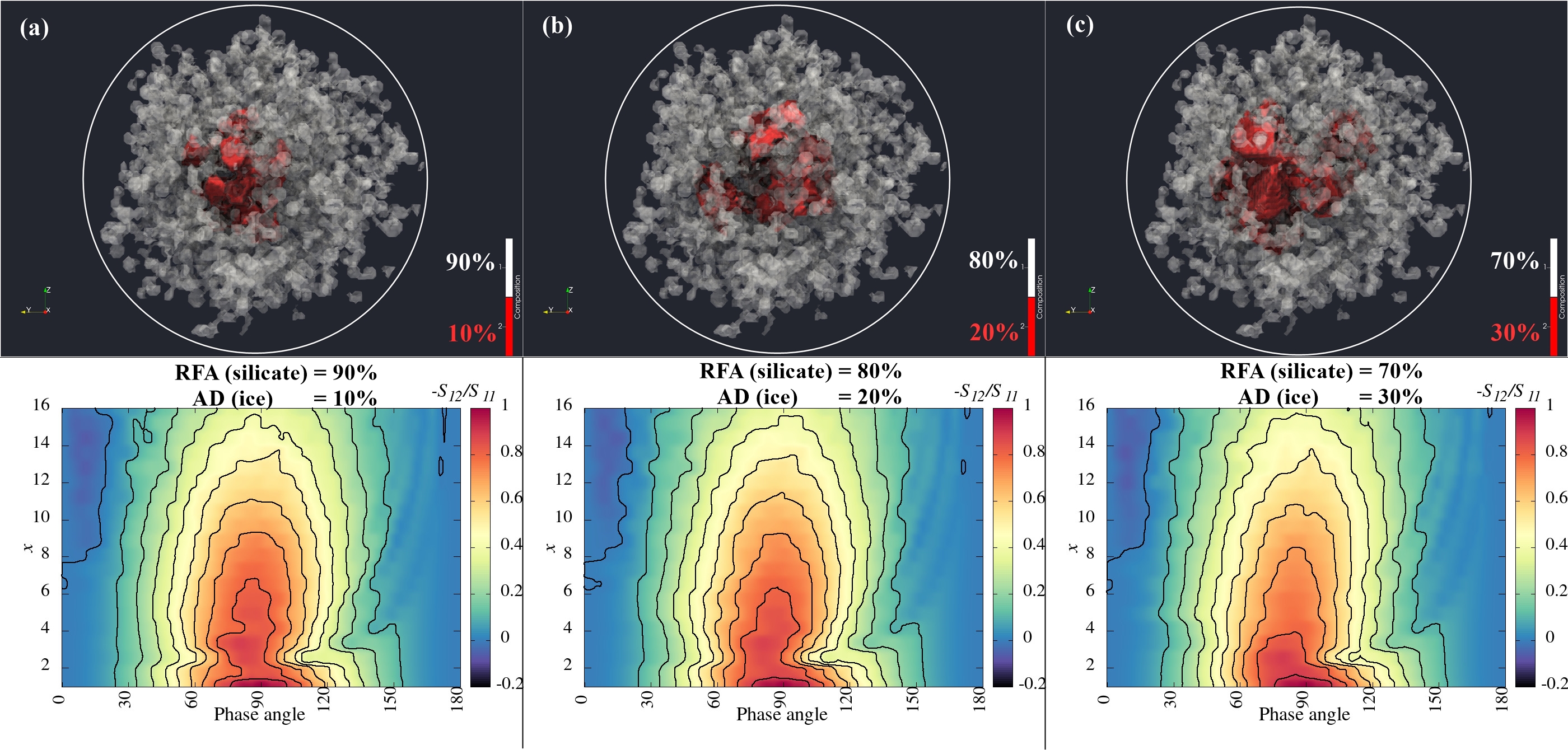}
    \caption{3D Visulization and respective variation in the degree of linear polarization with phase angle for $X$ = 2 to 16 over mixed morphology RFA-AD structures having mixing ratios (90:10) \emph{(a)}, (80:20) \emph{(b)} \& (70:30) \emph{(c)}}
    \label{fig:ff}
\end{figure*}

\begin{figure*}
    \centering
    \includegraphics[scale=0.65]{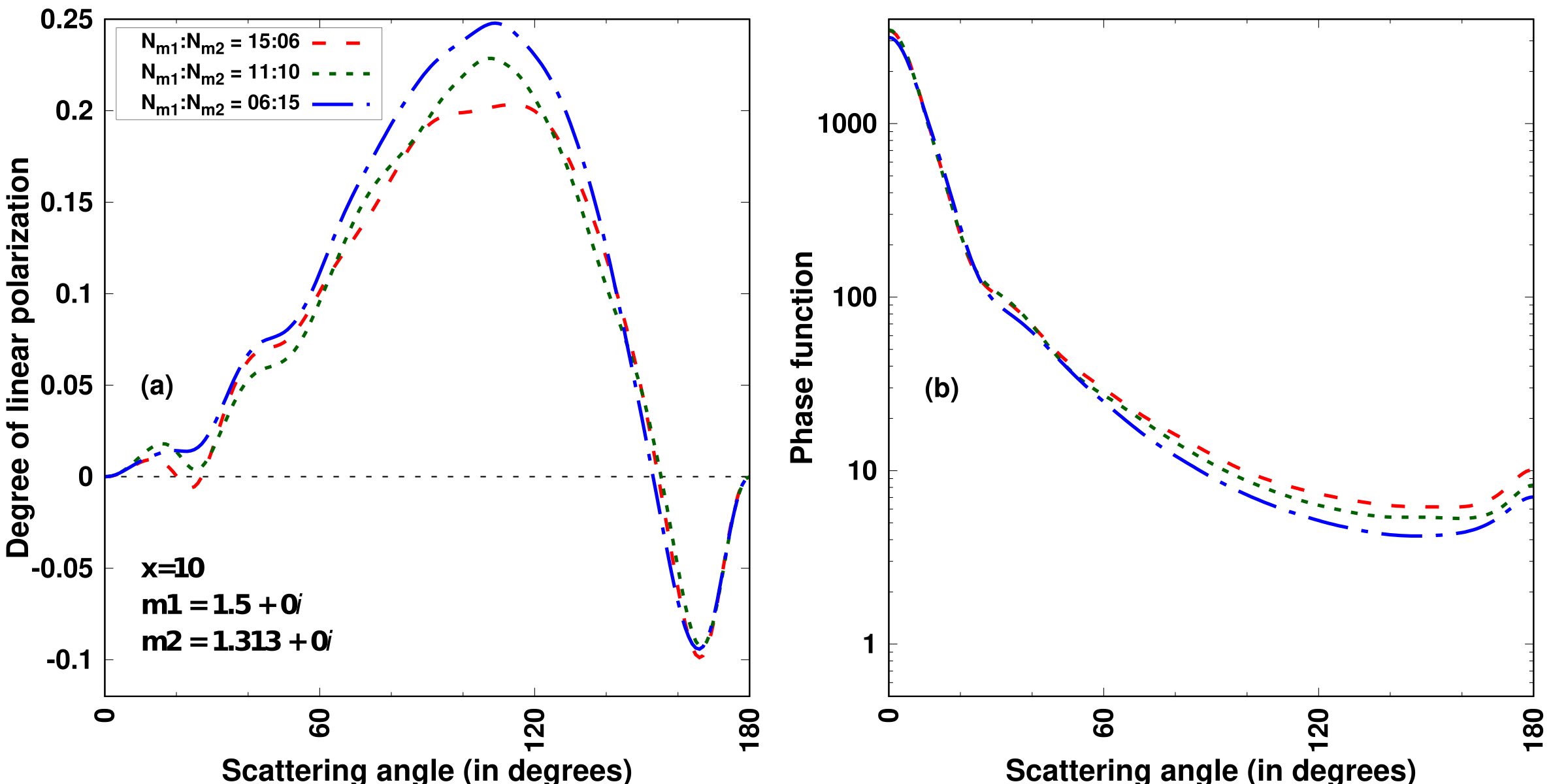}
    \caption{Verification of computational accuracy of the particles generated using REST: Recreating the variation of the degree of linear polarization (a) and phase function (b) with scattering angle for inhomogeneous AD particles following Figure-2 from \citep{Videen2015MixingScatterer}}
    \label{fig:videen_compare}
\end{figure*}

\subsection{Mixed Morphology (RFA-AD)}
As REST provides an opportunity to create numerous category of particles, in this section the development of new RFA-AD mixed morphology has been discussed. In various dusty regions such as cometary coma, protoplanetary disks and/or debris disks, there is high chance that particles of different morphologies may coexist as single entities. The significance of such complex morphologies are still unknown. Two structures RFA (high porous) and AD (low porous) having different morphologies are created using REST and an algorithm (not included in REST) is developed to fit the AD structures within the pores of the RFA structure. Further, three RFA-AD structures having RFA:AD ratios (90:10), (80:20) and (70:30) are created. Figures-\ref{fig:ff}(a-c) shows the variation of the degree of linear polarization ($-S_{12}/S_{11}$) with increasing phase angle ($0^{\circ} - 180^{\circ}$) over a size parameter ($x = 2{\pi}R/\lambda$) rang 2 to 16 for the three different mixing ratios of RFA:AD where the composition of RFA is silicate while that of AD is ice. 
 To verify the computational accuracy of particles generated using REST, Figure-2 of \cite{Videen2015MixingScatterer} is recreated, where the authors depict the degree of linear polarization and intensity obtained from light scattering simulations over inhomogeneous AD particles having a mixture of two different refractive indices for three different combinations as shown in Table-\ref{tab:table2}.
To generate inhomgeneous AD particles the PS and inhomogeneous algorithms (explained in Section-\ref{ps}) are considered. Finally, light scattering simulations are conducted for the three different combinations (15:6, 6:15, 11:10) of refractive indices (1.5+0$i$) \& (1.313+0$i$) for size parameter $x$ = 10 over 2000 different orientations. Figure-\ref{fig:videen_compare} shows that an increase in the amount of ice from 25\% - 75\%, the maximum polarization increases, the minimum polarization remains unchanged, while phase function ($S_{11}$) decreases.
\begin{table}
    \centering
    \begin{tabular}{|c|c|c|c|}
    \hline

       S/N & m1 & m2 & $N_{m1}$:$N_{m2}$ \\
    \hline
        1 & 1.5+0$i$ & 1.313+0$i$ & 15:6\\
        2 & 1.5+0$i$ & 1.313+0$i$ & 6:15\\
        3 & 1.5+0$i$ & 1.313+0$i$ & 11:10\\
    \hline
    \end{tabular}
    \caption{Three different combination of refractive indices used by \citeauthor{Videen2015MixingScatterer}(2015).}
    \label{tab:table2}
\end{table}

\section{Discussion \& Conclusions}
Following the growing research in the field of "Dust in Astrophysics" and the measurements of IDPs and comet dust from Earth-based and in-situ space missions, this study reports the development of a Java application package Rough Ellipsoid Structure Tools (REST). The application generates irregular dust structures that are constrained by physical parameters such as size, bulk-density, porosity, aspect-ratio and composition. These are the necessary parameters required for better interpretation of the light scattering and thermal emission observations from different dusty sources.
REST shall be very much useful to generate pristine cosmic dust particles found in comets, protoplanetary disks, debris disks and the interstellar medium. The morphology of RFA and AD structures generated using REST reassembles those from the original samples of IDPs and comets and shall provide better interpretations. Further, the presence of aspect-ratio in the super-ellipsoidal structures such as RS-SE, PS-SE and AD-SE, shall be useful to explain radiative-torque alignment in protoplanetary and circumstellar dust.
The flexibility of REST to craft structures by controlling the number of different seed cells, aspect ratio and exponents, provides a high possibility to create numerous type of dust morphologies. 
Practically, REST can be applied in various areas of research that incorporates material or dust properties. In astrophysics, it can be applied in the study of dust present in our Solar System (IDPs, comets, asteroids, Lunar dust,etc.). The rough super-ellipsoids generated using REST has crucial implications in the study of interstellar medium, where the dust particles are believed to have certain aspect ratio and are influenced by radiation pressure and galactic magnetic field.

\section*{Acknowledgements}
The author acknowledges the high-performance computing facilities (NOVA) of the Indian Institute of Astrophysics, Bangalore and Vikram-HPC of the Physical Research Laboratory, Ahmedabad, where all the intensive light scattering simulations are conducted. The author acknowledges Prof. Sujan Sengupta of IIA, Bangalore and Dr. Shashikiran Ganesh of PRL, Ahmedabad for important discussion. The author also acknowledges Prof. B.T. Draine for making DDSCAT publicly available.  

\newpage


\bibliography{references}{}
\bibliographystyle{aasjournal}



\end{document}